\documentclass{article}
\usepackage{hyperref}
\usepackage{graphicx}
\usepackage{comment}
\usepackage{amsmath,amssymb} % define this before the line numbering.
\usepackage{color}
 \usepackage{blindtext}
\usepackage[tableposition=top]{caption}
\usepackage{stackengine}
\usepackage{graphicx}
\usepackage{capt-of}
\usepackage{booktabs}
\usepackage{varwidth}
\usepackage{mathtools}
\usepackage{pdfpages}
\newsavebox\tmpbox
\DeclarePairedDelimiterX{\norm}[1]{\lVert}{\rVert}{#1}

\usepackage{floatrow}
\newfloatcommand{capbtabbox}{table}[][\FBwidth]

\usepackage[font=small,labelfont=bf,tableposition=top]{caption}

\DeclareCaptionLabelFormat{andtable}{#1~#2  \&  \tablename~\thetable}

\newcommand{\captionaboveof}[3][]{%
    \vskip-\abovecaptionskip
    \vskip+\belowcaptionskip
    \def\@captype{#2}%
    \ifx\@nnil#1\@nnil
        \caption{#3}%
    \else
        \caption[#1]{#3}%
    \fi
    \vskip+\abovecaptionskip
    \vskip-\belowcaptionskip
}
\makeatother

\usepackage{float}
\floatstyle{plaintop}
\restylefloat{table}

\usepackage{floatrow}
\usepackage{everypage}
\usepackage{afterpage}
\usepackage{ifthen}
\usepackage{caption}
\usepackage{subcaption}
\usepackage{float}
\usepackage{wrapfig}
\usepackage{graphicx}
\usepackage{subcaption}

\usepackage{ISBI_Template/spconf,graphicx}

\begin{document}
% \linenumbers
%\pagestyle{headings}
%\mainmatter
%\def\ECCVSubNumber{2873}  % Insert your submission number here

% \title{Shape-consistent Generative Adversarial Networks for multi-modal Medical segmentation maps}

% \name{}
% \address{}
% \name{Leo Segre*,\qquad Or Hirschorn*,\qquad Dvir Ginzburg,\qquad Dan Raviv}
% \address{leosegre@gmail.com\qquad or.hirschorn@gmail.com\qquad dvirginzburg@mail.tau.ac.il\qquad darav@tauex.tau.ac.il\\
% \\
% Tel Aviv University}
%\hspace{5pt}leosegre@gmail.com \hspace{20pt}or.hirschorn@gmail.com \\
%\hspace{5pt} dvirginzburg@mail.tau.ac.il \hspace{20pt}darav@tauex.tau.ac.il}

% \begin{comment}
% \titlerunning{Shape-consistent Generative Adversarial Networks for multi-modal Medical segmentation maps}
% \authorrunning{L. Segre and O. Hirschorn}
% \institute{Tel Aviv University}

\title{Shape-consistent Generative Adversarial Networks for multi-modal Medical segmentation maps}

% More than two addresses
% -----------------------
% \name{Author Name$^{\star \dagger}$ \qquad Author Name$^{\star}$ \qquad Author Name$^{\dagger}$}
%
% \address{$^{\star}$ Affiliation Number One \\
%     $^{\dagger}$}Affiliation Number Two

\name{Leo Segre\thanks{\textsuperscript{*}Equal contribution}\textsuperscript{*} \qquad Or Hirschorn\textsuperscript{*} \qquad Dvir Ginzburg \qquad Dan Raviv}
\address{Tel Aviv University}

%
%******************
\maketitle

%%%%%%%%% BODY TEXT
%-------- Abstract
\begin{abstract}
Image translation across domains for unpaired datasets has gained interest and great improvement lately. In medical imaging, there are multiple imaging modalities, with very different characteristics. Our goal is to use cross-modality adaptation between CT and MRI whole cardiac scans for semantic segmentation.
We present a segmentation network using synthesised cardiac volumes for extremely limited datasets.
Our solution is based on a 3D cross-modality generative adversarial network to share information between modalities and generate synthesized data using unpaired datasets. Our network utilizes semantic segmentation to improve generator shape consistency, thus creating more realistic synthesised volumes to be used when re-training the segmentation network.
We show that improved segmentation can be achieved on small datasets when using spatial augmentations to improve a generative adversarial network. These augmentations improve the generator capabilities, thus enhancing the performance of the Segmentor.
Using only 16 CT and 16 MRI cardiovascular volumes, improved results are shown over other segmentation methods while using the suggested architecture. Our code is publicly available\footnote{\href{https://github.com/orhir/3D-Shape-Consistent-GAN}{github.com/orhir/3D-Shape-Consistent-GAN}}.
\end{abstract}
%--------
%-------- Introduction
\section{Introduction}

%\begin{wrapfigure}{r}{0.5\textwidth}
%  \begin{center}
%    \includegraphics[width=\linewidth]{images/back_and_forth.png}
%  \end{center}
%  \caption{Self-supervised dense correspondence using a cycle mapping architecture. By minimizing the geodesic distortion only on the source shape, we can learn complex
%  deformations between structures.}
%  \label{fig:back_forth}
%\end{wrapfigure}

\begin{figure}[t]
\centering
\includegraphics[scale=0.395]{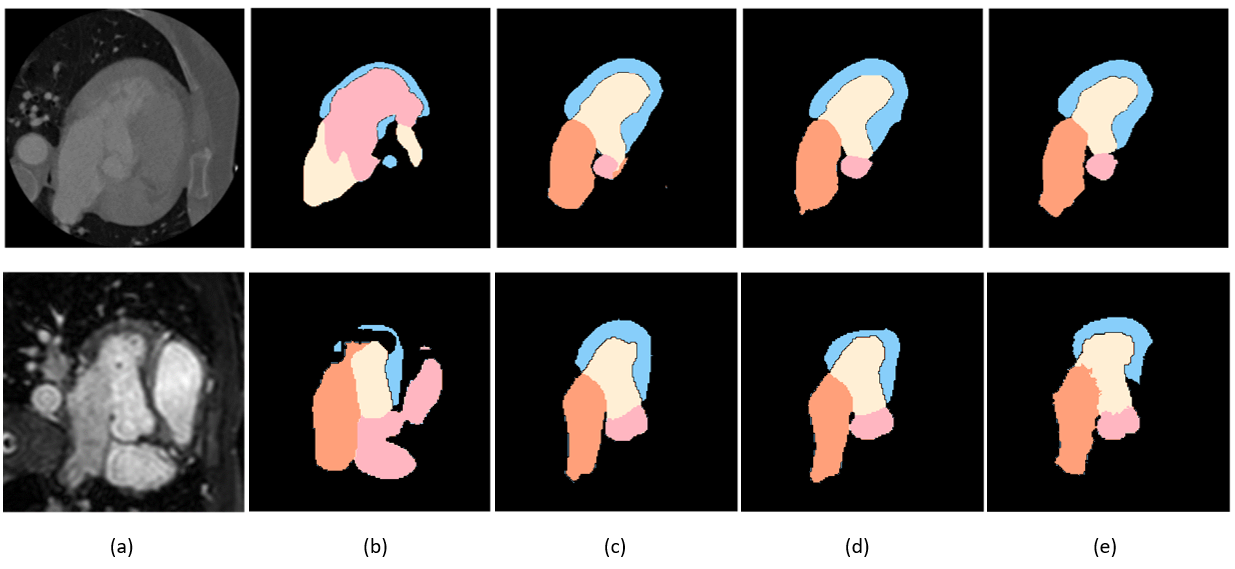}
\caption{Segmentation results generated by our method on cardiac CT images (top) and MRI (bottom). a) Examples of test images, b) Segmentation results without preprocess augmentations, c) Segmentation results without synthesized data, d) Segmentation results of our full method, e) The ground truth.}
\label{fig:segmentation}
\end{figure}

% Establishing the territory
Semantic segmentation is a key perceptual function in computer vision, aiming to densely categorize an image into meaningful distinguished areas. In the medical imaging domain, semantic segmentation is vital, providing tools for diagnostics, treatment planning, and prognosis. For disease diagnostics and surgical needs, multiple imaging modalities are available such as MRI, CT, and X-ray. \\
Traditional machine learning methods, such as atlas and model-based methods, showed good performance in cardiac image segmentation. However, they usually require features engineering which differs between image modalities~\cite{10.3389/fcvm.2020.00025}. In contrast, deep learning algorithms show promising results while implicitly discovering features from the data. They have been widely adopted for various tasks, from 2D binary segmentation~\cite{zhang2021transfuse} to multi-class 3D segmentation ~\cite{10.1007/978-3-319-46723-8_49}.\\
% Establishing a niche
However, to achieve satisfying results, a sufficiently large number of training samples is required. As demonstrated in figure~\ref{fig:segmentation}, the results of semantic segmentation when the training set is too small are significantly degraded. Segmented medical imaging datasets are hard to acquire and use due to strict regulations, lack of support from hospitals in acquiring the data, and high costs of medical imaging services. As medical images acquired using different modalities have very different characteristics, it is especially challenging to obtain data for new imaging modalities. This also applies to images from the same modality, captured using different scanning machines. \\
To overcome the lack of data, a common way to generate synthesized data is to use augmentations on the original data~\cite{P_rez_Garc_a_2021}. 
%%%%%%%%%%%% Needed?? %%%%%%%%%%%%
% While 2D augmentations are computational efficient, 3D augmentations are shape consistence but computational hard.
% When segmenting 2D slices it is possible to apply augmentations online, but when using 3D online augmentations it appears to be time consuming in inefficient way. 
% Hence, 3D voxels augmentations should be a part of the preprocess phase and they perform better than 2D augmentations. Since the task is segmentation, each pixel has its own % label. It is important to maintain the correct label for each pixel by applying the exact same spacial augmentations on the label map and the original data. \\
%%%%%%%%%%%%%%%%%%%%%%%%%%%%%%%%%%%%%%%% 
Another way to generate synthesized data is using a Generative Adversarial Network (GAN). This can be done directly by separate GAN for each domain, but it is even more effective to use cross-modality GAN and share information between modalities and generate more accurate synthesized data~\cite{chen2020unsupervised,zhang2019translating}. Although some models used GANs as a proxy task to the segmentation objective on very small datasets~\cite{chen2020unsupervised}, the mentioned method used 2D slices instead of 3D volumes thus discarding crucial information regarding the spatial consistency in the Z-axis.
%Since our data is divided into two unpaired domains, we used CycleGAN~\cite{zhu2020unpaired} concept to perform cross-domain translation. That method can provide more labeled data under the assumption that the source and target represent the same organ in different domains.\\
 % Occupying the niche
This paper presents a method to use deep learning based segmentation on a limited dataset by using augmentations and generating 3D synthesized shape-consistent data. The presented method provides significant improvements in the limited training data domain, with an average accuracy increase of 15.9\% in the final segmentation score. Additionally, we provide our code for future works.
%You need to have a contributions subsection in the introduction, where you outline the things are unique to your method.

%-------- Related Work
\section{Related Work}
Using deep learning for domain adaptation in a supervised or unsupervised manner has recently gained popularity~\cite{chen2020unsupervised, 7410436, liu2018unsupervised, kamnitsas2016unsupervised}. As the availability of training data from the same set of subjects in both source and target modalities is undesirable (requires multiple scans from each subject), an unsupervised cross-modal image synthesis, without pairing training data approach is beneficial~\cite{7410436}.\\
Since the rise of deep learning, image to image translation is usually formulated as a pixel to pixel mapping using CNN's (Convolutional Neural Network) encoders and decoders~\cite{zhang2019translating, isola2018imagetoimage, liu2018unsupervised}. CycleGAN had wide success, where bi-directional image translations are learned by two GANs separately, and the consistency constraint between transforms is enforced to preserve semantic information between transformed outputs~\cite{zhu2020unpaired}. The good results and robustness Cycle-GAN showed for many applications made it a popular backbone in future works. Thus, many unpaired image to image transformations are based on this framework with additional constraints to further regularize the transformation process. Although CycleGAN was initially proposed for 2D images, GANs have also been applied in 3D~\cite{abramian2019generating}.\\
For medical image processing, adversarial learning has presented great efficiency on a variety of tasks~\cite{inbook2018}.
Using synthetic data as augmented training data helps the segmentation network as seen for brain MRI and CT images~\cite{kamnitsas2016unsupervised}. \textit{Zizhao Zhang et al.}~\cite{zhang2019translating} used a segmentation network to force shape consistency of the output transformed domain, using a private massive data set. As massive datasets are usually hard to acquire in this domain, our work avoids this demand, presenting great results using a public dataset with only 20 CT and 20 MRI samples. 
SIFA~\cite{chen2020unsupervised} used a novel approach of fusing feature and image appearance adaptation and applied it to cross-modality segmentation of cardiac volumes, but was limited to 2D slices, losing z-axis shape data.
% As the use of augmentations was essential, and processing of 3D medical images such as MRI or CT presents different challenges compared to RGB images, \textit{Perez Garcia et al.}~\cite{perez-garcia_torchio_2021} suggests an open-source python library for preprocessing and augmenting medical images.\\
% Other methods without train set limitation of 16 scans per domain achieved great results on the full test set of MMWHS challenge 2017 ~\cite{Zhuang2019} and on large unpublished dataset~\cite{zhang2019translating}.
%-------- Implementation
\section{Methods}
\begin{figure*}[t]
\centering
\includegraphics[scale=0.65]{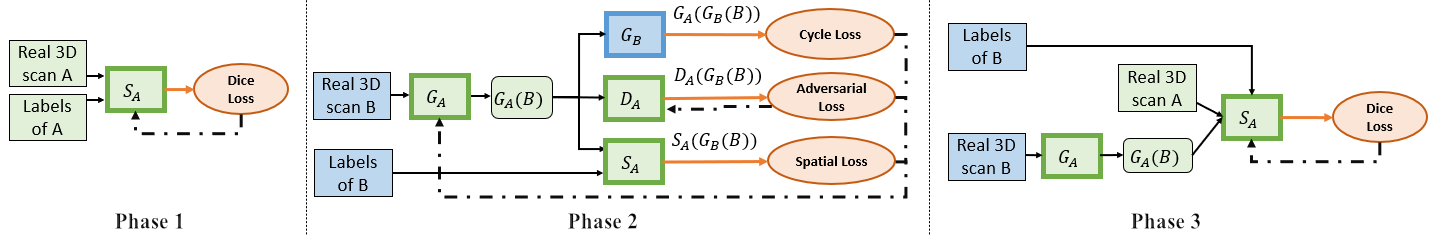}
\caption{Overview of our domain adaptation and segmentation architecture. The architecture above is duplicated to handle both domains, a duplication where "A" is CT and "B" is MRI and vise versa. The flow is based on three chronological phases as described in training strategies.}
\label{fig:arch2}
\end{figure*}

The method is applied in three phases. First, a segmentor is trained with the original data. Then, a 3D cycle and shape consistent GAN network is trained to create synthesized cardiac volumes, which is used later to train the segmentor in phase 3 and achieve the goal of improving the 3D segmentor network. Figure ~\ref{fig:arch2} illustrates the proposed architecture for domain adaptation in cardiac volumes. 

%-------- --------  --------  --------  --------  --------  --------  --------  --------  --------  -------- 
\subsection{Synthetic Data Generation}
First, all scans aligned to RAS+ orientation and following \textit{Cheng Chen et al.}~\cite{chen2020unsupervised} we manually cropped the MRI scans around the heart. To align between the CT scans and MRI scans, we also resized the CT scans into 256x256xZ, where Z preserves the original scan width-depth ratio.
Then, using TorchIO~\cite{P_rez_Garc_a_2021} we randomly applied numerous 3D transformations on both the cardiac volumes and labels, specifically designed for medical data: normalization, random anisotropy - Simulates an image that has been acquired using anisotropic spacing and resampled back to its original spacing, random elastic deformation - A random displacement according to a grid of control points, random affine - applies affine transformation and resamples back to the original spacing. 
% As the use of augmentations was essential, and processing of 3D medical images such as MRI or CT presents different challenges compared to RGB images, \textit{Perez Garcia et al.}~\cite{perez-garcia_torchio_2021} suggests an open-source python library for preprocessing and augmenting medical images.\\
%We also cropped the cardiac volume according to the labeled date during training to decrease the data input size as 3D convolutions are highly memory consuming.
We observed the best results for 200 3D synthetic augmented scans, as more synthetic scans resulted in significantly longer run times and negligible improvements.
%-------- --------  --------  --------  --------  --------  --------  --------  --------  --------  -------- 

\subsection{Volume to Volume Translation}
For two unpaired domains A and B, we employ generative adversarial networks using a generator $G$ and a discriminator $D$ for each domain. The volume to volume translation is based on CycleGAN U-Net architecture enhanced to 3D volumes using 3D convolutions.
The generator transforms the input domain A to the other domain B, the notation for this transformation is $G_B(X_A)$. The discriminator competes with the generator, trying to distinguish between a fake volume $G_B(X_A)$ and a real volume $x_B$. L2 loss is used to minimize the generator's objective of creating realistic volumes, noted as $\mathcal{L}_{adv, A}$.

%\[ \mathcal{L}_{adv, A} = 1-(prediction-target)^2 \]

%where the prediction is the discriminator's output for the generated volume and the target is 1.\\

We adopt the CycleGAN's approach of cycle consistency loss. Thus, we force the generators to reconstructed synthesized volumes $G_A(G_B(x_A))$ and $x_A$ to be identical. We encourage the transformed volumes to preserve content from the original volume using L1 loss:

\[ \mathcal{L}_{cycle, A} = -\frac{1}{N} \sum_{i} |x_A-G_A(G_B(x_A))| \]

Using the above constraints can lead to geometrically distorted transformations. The cycle consistency loss isn't enough to prevent spatial distortions. It is possible for generator B to create a distortion $F$ and for generator A to apply the reverse transformation $F^{-1}$ leading back to the original shape. Thus, a shape consistent constraint is needed to reduce the spatial distortion.\\
We suggest using a 3D segmentor to preserve shape consistency. The segmentor maps $x_i$ $\rightarrow$ $Y$ , where i is the domain A or B. To constrain the geometric invariance of the generated volume we optimize the Cross Entropy+Dice loss~\cite{isensee2018nnunet} of the generated domain and its labels:
\[ \mathcal{L}_{spatial, A} =  \mathcal{L}_{CE, A} + \mathcal{L}_{DICE, A}\]
\[ \mathcal{L}_{DICE, A} = -\frac{2}{|K|}\sum_{k\in K} \frac{\sum_{i\in I}{u_{i}^{k}}{v_{i}^{k}}}{\sum_{i\in I}{u_{i}^{k}}+\sum_{i\in I}{v_{i}^{k}}}\]
where u is the softmax output of the network and v is a one hot encoding of the ground-truth segmentation $y_A$ of the volume $x_A$.
%\[ \mathcal{L}_{spatial, A} = -\frac{1}{N} \sum_{i} y_{A,i} log(S_B(G_B(X_A))_{i,y_i}) \]
% where $y_A$ denotes the true label of the volume $x_A$.
The generator's total loss function is composed of all the above constraints:
\[ \mathcal{L}_{A} =  \lambda_{adv}\mathcal{L}_{adv, A} + \lambda_{cycle}\mathcal{L}_{cycle, A} + \lambda_{spatial}\mathcal{L}_{spatial, A}\]
where $\lambda_{i}$ is a trade-off parameter.

%-------- --------  --------  --------  --------  --------  --------  --------  --------  --------  -------- 
\subsection{Segmentation}

The segmentation network in our solution is based on a 3D U-net architecture~\cite{10.1007/978-3-319-46723-8_49}. In our architecture, there are two identical segmentors, one for each domain. Although each segmentor has its own source domain, both CT and MRI segmentors have the same target domain - cardiac segmentation labels. Thus, the segmentation task can be implemented on the original volume or on a generated synthetic volume, in both cases the ground-truth labels are identical. Formally, given input voxel $X_A$ from domain A and its labels map $Y_A$, we define the following. 

\[ \mathcal S_A(X_A) = S_B(G_B(X_A)) = Y_A \]

Using this approach, we can train the segmentor with either real and synthesised data, it is important since segmentation networks usually require a lot of data to be trained on. Hence, the losses of the segmentors are defined by $L_{seg}$ and $L_{seg,syn}$. Both are Cross Entropy+Dice losses, as used in \textit{Fabian Isensee et al.}.

\[ \mathcal L_{seg}(X_A, Y_A, S_A) = \mathcal{L}_{spatial}(Y_A, S_A)\]
\[ \mathcal L_{seg,syn}(X_A, Y_A, S_B) = \mathcal{L}_{spatial}(Y_A, S_B(G_B(X_A)))\]

%\[ \mathcal L_{seg}(X_A, Y_A, S_A) = -\frac{1}{N} \sum_{i} Y_i log(S_A(X_A)_{i,Y_i}) \]
%\[ \mathcal L_{seg,syn}(X_A, Y_A, S_B) = -\frac{1}{N} \sum_{i} Y_i log(S_B(G_B(X_A))_{i,Y_i}) \]

%%%%%%%%%%%%%%%%%%%%%%%%%%%%%%%%  NEEDED ????   %%%%%%%%%%%%%%%%%%%%%%%%%%%%%%%%  
For each voxel of the input, the segmentor evaluates a vector of probabilities - one for each label. $S_A(X_A)_{i,Y_i}$ denotes the output of $S_A(X_A)$ on voxel i regarding to the probability of the ground-truth label of this voxel.
$argmax(S_A(X_A)_i)$ will provide the label prediction of voxel i.
%%%%%%%%%%%%%%%%%%%%%%%%%%%%%%%%%%%%%%%%%%%%%%%%%%%%%%%%%%%%%%%%%%%%%%%%%

%-------- --------  --------  --------  --------  --------  --------  --------  --------  --------  -------- 
\subsection{Training Strategies}

Training the network consists of three phases, as we observed pre-training and fine-tuning the segmentor and generator results in better performances. \\
First, the segmentor is pre-trained for 100 epochs using only the original data. %During data load, we crop the 3D volume according to the input labels, thus removing mainly background volume. This way we can decrease the size of the input images (as 3D CNN networks are highly memory consuming) while minimizing the loss of valuable data.
Then the generator and discriminator are trained leaning on the pre-trained segmentor from phase 1. The goal of this phase is to train a generator that can generate synthesized data in addition to the preprocessed synthesized volumes. The generator and discriminator are trained for 50 epochs without the spatial loss, and then for another 150 epochs with the spatial loss enforcing shape consistency.
Last, after having a trained shape consistent generator, we train the segmentation network using the augmented and synthesised 3D volumes for 100 epochs.

%-------- --------  --------  --------  --------  --------  --------  --------  --------  --------  -------- 
\subsection{Network Configurations and Implementations}

%%Our code is publicly available at \href{https://github.com/orhir/hape-consistent-Generative-Adversarial-Networks-for-multi-modal-Medical-segmentation-maps}{github.com/orhir/hape-consistent-Generative-Adversarial-Networks-for-multi-modal-Medical-segmentation-maps}.\\
Our network consists of segmentor, generator and discriminator modules for each domain.\\
The segmentation network is a 3D U-net consisted of 4 downsampling convolutions (maximum downsample rate is 16) and 4 upsampling using nearest interpolation, with convolutions of 3x3x3 kernel and stride 1. \\
The discriminators follow PatchGAN configurations~\cite{9157688}, consists of 3 convolutional layers with kernels 4x4x4 and stride 2, and 2 convolutional layers with stride 1. For the first 4 layers, each convolutional layer is followed by a normalization layer and leaky ReLU with 0.2 slope parameter. \\
The generator is based on CycleGAN's U-net, adjusted to 3D volumes, using a skip-connection U-net, as it achieves faster convergence and locally smooth results~\cite{zhang2019translating}. We apply 5 downsampling with 3x3x3 kernel and stride 2, and upsample using nearest interpolation with 3x3x3 kernel and stride 1.
We implemented our framework in PyTorch, and the training was done on 8 NVIDIA Quadro RTX 8000 GPUs. All the networks were optimized using the Adam optimizer with a learning rate of $2\times10^{-4}$.

%-------- Results
\section{Results}

\begin{table}
\centering

\begin{tabular}{lcc}

\toprule

\textbf{Model}  &\textbf{CT} &\textbf{MRI} \\
\midrule
 
SynSeg-Net
&49.7 &  58.2\\

AdaOutput
&51.9 & 59.9\\

PnP-AdaNet
&54.3 & 63.9\\
 
SIFA
&63.4 & 74.1\\

Ours - 4 labels
&\textbf{88.2} & 81.2\\

Ours - 7 labels
&85.0 &\textbf{81.8}\\

\bottomrule
\end{tabular}

\caption{Segmentation performance comparison. The fifth and sixth rows show the boosted results by using shape consistent synthetic data, comparing SynSeg-Net, AdaOutput, PnP-AdaNet, SIFA
and our method (using 4 labels and 7 labels), respectively.
}
\label{Tab:diceScore}
\end{table} 
We use the Multi-Modality Whole Heart Segmentation (MMWHS) Challenge 2017 dataset for cardiac segmentation \cite{ZHUANG201677, fully_auto_whs}. This dataset consists of unpaired 20 CT and 20 MRI volumes with 7 segmented labels. We divided the dataset as commonly used to 80\% training data and 20\% test data. As \textit{Cheng Chen et al.} is a major leading paper in this field and particularly on this dataset, we aligned our test and train samples to it and followed its test protocol for a better comparison. To evaluate the performance of the network segmentation accuracy we employ the commonly-used Dice similarity coefficient. As in previous works, the score is an unweighted average of the labels dice score, where each label's dice score is calculated separately.
We compare our method with the SOTA unsupervised domain adaptation methods which utilize either feature alignment, image alignment, or their mixtures as shown in table~\ref{Tab:diceScore}. As part of our method we generated total of 400 3D augmented scans, other methods provided with a total of 21,600 2D augmented slices prepossessed as in \cite{chen2020unsupervised}. Although the original dataset contains 7 labels for each volume, earlier works aimed to segment only 4 labels: ascending aorta (AA), left atrium blood cavity (LAC), left ventricle blood cavity (LVC), and myocardium of the left ventricle (MYO). Thus, we first trained our network to segment only those 4 labels and compared the results to previous unsupervised networks. All four previous compared models are using 2D slices for the segmentation and as can be seen in table~\ref{Tab:diceScore}, SIFA achieved on this dataset 63.4\% CT and 74.1\% MRI dice score. Our method which is the only method in table~\ref{Tab:diceScore} that uses a 3D segmentor achieves 88.2\% CT and 81.2\% MRI dice score. This shows empirically that using 3D convolutions has an implicit effect of smoothness and z-coherency between the different slices.\\
As a second step, we aimed to segment all 7 labels and test the results compared to the 4 labels segmentation of our network. In addition to the first 4 labels, we also segmented: right atrium blood cavity (RAC), right ventricle blood cavity (RVC), pulmonary artery (PA). 
%%%%%%%%%%%%%%
Our method achieves 85.0\% CT and 81.8\% MRI dice score. CT dice score is slightly decreased compared to 4 labels segmentation(3.6\%). The MRI dice score slightly increased.
The task of segmenting all 7 labels is more valuable since segmenting only 4 labels is not practical for real-life applications. Thus, our method shows promising results in segmenting multiple labeled volumes.\\
%%%%%%%%%%%%%%%%%
To observe the effectiveness of our suggested architecture we conducted ablation experiments as shown in table~\ref{Tab:ablation}. First, we trained the network on the original dataset without preprocessed augmentations, the score of both CT and MRI are very low compared to the full method. This is expected since the original dataset is extremely small, but it also proves the effectiveness and importance of our augmentations. Another ablation study is to completely leave out the synthesized data and train the segmentor using only the preprocessed data. It stands out that there is a major difference between CT and MRI results in this case, While the CT domain performed a 7.2\% improvement when used the synthesised data, the MRI domain shows the best result without any generated data at all. A possible explanation is that the MRI scans in this dataset are more diverse than the CT scans. Where in the MRI domain some scans are originally oriented to (P,S,R) and other scans to (L,S,P) and the size differs between scans, the CT scans are homogeneous in terms of orientation and size. We aim to better understand how to generate MRI synthesized scans in future studies. In the last ablation study, we did not use any shape consistency loss, which means we generated data without any segmentation information. It can be seen that the shape consistency constraint improves the score in both domains.
\begin{table}
\centering

\begin{tabular}{lcc}
\toprule

\textbf{Mode}  & \textbf{CT} & \textbf{MRI}

\\
\midrule
Preprocessed Synthesized Volumes
&58.7 & 61.7\\

Cross-Domain Synthesized Volumes
&81.0 & \textbf{81.9}\\
 
Shape Consistency
&86.2 & 80.5\\
  
Full method
&\textbf{88.2} & 81.2\\
 
\bottomrule
\end{tabular}
\caption{Ablation study on our suggested net evaluating F1 score. The "Mode" column states the switched \textbf{off} component.
}
\label{Tab:ablation}
\end{table} 
%-------- Conclusion
\section{Conclusion}
We have presented a method for Whole Heart Segmentation from a limited dataset using augmentations and generated shape consistent synthetic data. Our results on the MICCAI
2017 Multi-Modality Whole Heart Segmentation challenge show excellent performance for CT scans. 
As can be observed, using 3D segmentation rather than 2D segmentation on each Z-axis slice, achieve boosted results on both modalities. This is empirical proof of the added z-axis shape data. \\
%-------- Ethical
\section{COMPLIANCE WITH ETHICAL STANDARDS}
This research study was conducted retrospectively using human subject data made available in open access by MMWH Challenge 2017. Ethical approval was not required as confirmed by the license attached with the open access data.

%-------- Acknowledgments
\section{acknowledgments}
This work is partially funded by the Zimin Institute for Engineering Solutions Advancing Better Lives, the Israeli consortiums for soft robotics and autonomous driving, the Nicholas and Elizabeth Slezak Super Center for Cardiac Research and Biomedical Engineering at Tel Aviv University and TAU Science Data and AI Center. 

\bibliographystyle{ISBI_Template/IEEEbib}
\bibliography{egbib}

%\cite{SGMDS_axiomatic,halimi2019unsupervised,functional_cor,rodola2014dense}
%\cite{fmnet,3d_coded,halimi2019unsupervised,Boscaini2016AnisotropicDD,masci2015geodesic,litman2013learning}
%\linebreak\linebreak
%$\mbox{\Large\( shared\  weights \)}$

\clearpage

\end{document}